# Content-Aware Foveated Camera for Multi-Target Tracking

Zihan Zang[†], Do Young Kim[†], Yifeng Zeng, and Liang Gao[*]

*Department of Bioengineering, University of California, Los Angeles, Los Angeles, California 90095, USA*
[†]*These authors contributed equally.*
[*]*gaol@ucla.edu*

**Abstract:** Modern image sensors deliver substantial space-time bandwidth, yet indiscriminate acquisition often overwhelms memory, computation, and downstream perception. We present a content-aware, multi-foveated camera that dynamically reallocates sensing and magnification to multiple regions of interest (ROIs). A phase-only spatial light modulator (SLM) serves as a solid-state, inertia-free beam-steering and lens element, enabling per-frame field-of-view (FOV) reconfiguration and content-aware target tracking. By interleaving frames across foveae, our system preserves a wide-FOV situational context while refreshing each ROI at high rates, thereby reducing data volume without degrading task performance. We constructed a prototype employing single-SLM, single-sensor architecture and demonstrated its application in real-time multi-object tracking with dynamic ROI maintenance across multiple viewpoints. The approach offers a general pathway to integrate detection, tracking, and segmentation algorithms in the acquisition loop, shifting workload from post hoc processing to intelligent capture.

## 1. Introduction

The rapid growth of imaging technologies has created a widening gap between data acquisition and analysis. Modern cameras can capture gigapixel frames at video rates, advanced 3D microscopes produce terabytes of data per hour, and autonomous imaging platforms routinely generate hundreds of gigabytes to tens of terabytes of data per hour [1-3]. This explosive growth in data generation far outpaces advances in computational processing, resulting in a fundamental bottleneck between image capture and analysis [4].

Biological vision suggests a different approach: allocate high resolution only where it is most needed while preserving wide-field context. Foveated camera designs adopt this principle using reconfigurable optics that concentrate sampling on task-relevant regions of interest (ROIs). Mechanical implementations achieve this by laterally shifting lens assemblies to scan the field of view (FOV) [6]; rotating Risley prisms can accelerate scanning and have been applied to foveated imaging, though their fixed trajectories limit flexibility [7]. In contrast, fast micromirror systems offer far greater agility, rapidly revisiting multiple ROIs at video rates for multi-target tracking while maintaining a broad FOV [8]. In mesoscopy, random-access widefield approaches divide the scene into sub-FOVs and scan them sequentially, enabling centimeter-scale coverage with subcellular resolution, yet they suffer from the inertia and settling delays inherent to mechanical scanners [9].

Solid-state deflectors eliminate mechanical inertia. Acousto-optic deflectors (AODs) support microsecond-scale, random-access point visits, and they have long underpinned random-access two-photon scanning; however, their strengths lie in point-scan sampling, rather than rapid, whole-image reconfiguration [10]. Digital micromirror devices (DMDs) enable very high-speed structured illumination and holography, yet their binary amplitude nature forces phase encoding with reduced diffraction efficiency, which makes full-frame wavefront steering less efficient than phase modulation [11,12]. By contrast, liquid-crystal-on-silicon spatial light modulators (LCOS-SLMs) provide true phase control with millisecond-scale response that is largely independent of steering angle, enabling per-frame changes in focus, zoom, tilt, and higher-order corrections without trajectory planning [13]. Early foveated imaging with phase SLMs focused on reconfigurable aberration correction [14] and dual-fovea variants [15], while related work explored line-sensor systems using SLM-generated complex wavefronts [16]. Recent microscopy systems pair linear SLMs with custom beam rotators for random-access point scanning and combine SLMs with microlens arrays for optical segmentation [17,18]. Despite these advances, existing foveated systems remain limited: they cannot dynamically adapt ROI selection based on the scene content, and they underutilize the intrinsic solid-state advantage of deterministic, uniform switching across deflection angles.

Here, we introduce a content-aware foveated camera that uses a phase SLM as a programmable lens to fast-steer multiple ROIs on demand, while preserving a wide-FOV context on a conventional sensor. By embedding content-aware algorithms (such as object detection and tracking) inside the acquisition loop, the camera reallocates spatiotemporal sampling to ROIs with the highest informational relevance. Importantly, the SLM's solid-state, inertia-free response permits per-frame FOV reconfiguration without path planning or settling penalties that depend on deflection amplitude, enabling simple, deterministic "frame-interleaved" servicing of multiple foveae. In practice, the sensor alternates frames among ROIs, maintaining high update rates for each target while periodically refreshing the context view. We prototyped the system with a single SLM and a commercial image sensor, demonstrating video-rate multi-object tracking with dynamic ROI maintenance. We showcased that task-relevant performance can be preserved while substantially reducing raw data throughput. The proposed content-aware foveated paradigm outlines a general passway from passive imaging to closed-loop perception.

## 2. Principle

*2.1 SLM-synthesized diffractive lens*

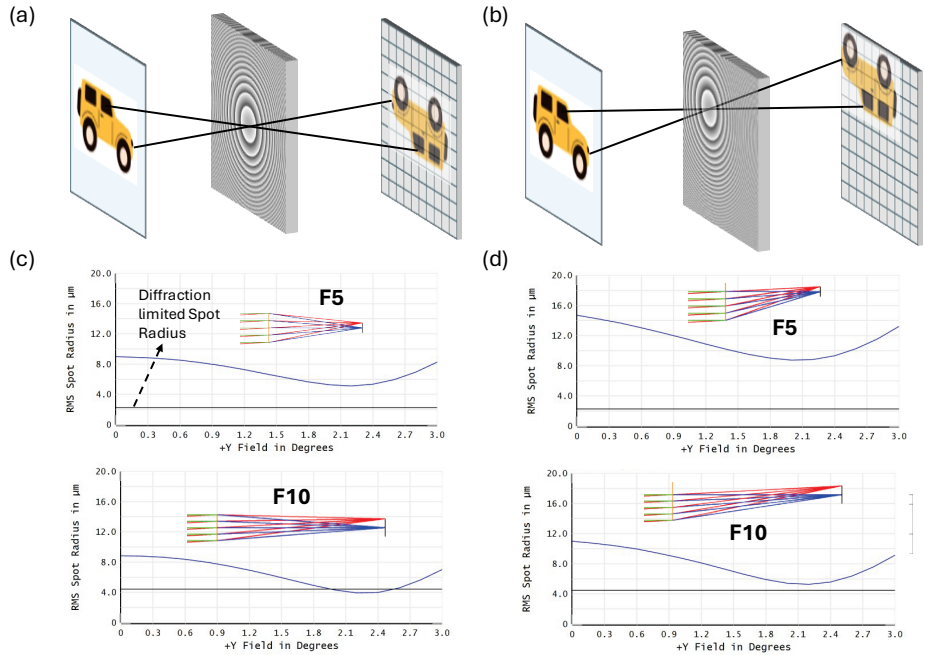

Fig. 1. SLM-synthesized diffractive lens. (a) on-axis configuration; (b) Off-axis configuration; (c, d) RMS spot radius versus field angle for (a) and (b) with F-number = 5 and 10; $\lambda$ = 550 nm and aperture = 10 mm. Phase profiles are optimized to minimize RMS spot size across the field.

A phase spatial light modulator based on LCOS can update at millisecond—or faster—timescales while displaying arbitrary, spatially varying phase patterns. This capability enables the SLM to synthesize a diffractive (Fresnel) lens that produces diffraction-limited focusing for a normally incident monochromatic plane wave. Let $(x, y)$ denote coordinates across an SLM with an aperture of diameter $D$, $r = \sqrt{x^2 + y^2}$, and wavelength $\lambda$. The exact spherical-wave phase that yields on-axis focus at distance $f$ is, up to an irrelevant constant and modulo $2\pi$,

$$\phi_0(x, y; f) = -\frac{2\pi}{\lambda}\left(\sqrt{f^2 + r^2} - f\right) \pmod{2\pi}, \quad (1)$$

which reduces in the paraxial limit $r \ll f$ to the familiar quadratic Fresnel lens $\phi_{\text{parax}}(x, y; f) \approx -\frac{\pi}{\lambda f}r^2$. Using the exact square-root form ensures high-fidelity focusing even for relatively large apertures. Off-axis focusing is obtained by shifting the center of curvature of the converging spherical wave, equivalently adding a linear tilt to the Fresnel lens. If the desired focal point at the focal surface is laterally displaced by $(x_f, y_f)$, the exact off-axis phase can be written as

$$\phi(x, y; f, x_f, y_f) = \frac{2\pi}{\lambda}\left(f - \sqrt{f^2 + (x - x_f)^2 + (y - y_f)^2}\right) \pmod{2\pi}. \quad (2)$$

For small angles, $(x_f, y_f) \approx f(\theta_x, \theta_y)$, so steering by angles $(\theta_x, \theta_y)$ can be viewed either as a lateral shift of the focus in the image plane or as an added linear phase gradient on the pupil. Translating the underlying Fresnel phase in this way realizes an off-axis diffractive lens, as conceptually illustrated in Fig. 1(a, b).

When used as an imaging element, a Fresnel lens implemented on the SLM behaves, over a controlled field of view, like a thin lens for monochromatic light. If an object at a plane at distance $z_o$ is to be imaged to a sensor plane at $z_i$, the Gaussian imaging relation

$$\frac{1}{f} = \frac{1}{z_o} + \frac{1}{z_i}, \quad m = -\frac{z_i}{z_o} \quad (3)$$

sets the conjugates and magnification $m$. The on-axis diffraction-limited point-spread function (PSF) is governed by the pupil F-number $FN = f/D$; the Airy disk diameter to the first zero at the image plane is $2.44\lambda \cdot FN$. Beyond the paraxial regime, geometric aberrations arise from the diffractive lens structure. Ray tracing reveals these aberrations, which can be minimized by modifying the phase profile in Eq. 2. This optimization balances aberration performance across the full field of view for given lens specifications: F-number, FOV coverage, and off-axis angles $(\theta_x, \theta_y)$.

Figure 1(c) plots the root-mean-square (RMS) spot radius versus field angle for FN = 5 and FN = 10; Figure 1(d) shows the corresponding results for an off-axis configuration in which the optical axis passes through the edge of the clear aperture. All simulations assume $\lambda$ = 550 nm and a 10 mm aperture. For each configuration, the lens phase profile is re-optimized to minimize RMS spot radius across $\pm 3°$ FOV, which improves imaging quality compared to the direct use of Eq. 2. During optimization, we augment the ideal off-axis Fresnel phase with a radially symmetric even-order polynomial in the local radius $r$ about $(x_f, y_f)$:

$$\phi_{\text{opt}}(x, y) = \phi(x, y) + \sum_{m=1}^{n} A_i r^{2m}, \quad (4)$$

where $r^2 \equiv (x - x_f)^2 + (y - y_f)^2$. The phase is treated as continuous (unwrapped) during optimization and wrapped modulo $2\pi$ after optimization. Under moderate steering angles and FOVs, the predicted PSF remains close to the diffraction limit for both on-axis and off-axis operations, indicating that high-quality monochromatic imaging is achievable when the F-number and FOV are properly chosen. If desired, the correction can be parameterized with rotationally asymmetric terms, including both shifted odd-order and cross terms, or equivalently with shifted Zernike modes of nonzero azimuthal order to target specific off-axis aberrations.

*2.2 Content-Aware Foveated Imaging via Frame Interleaving*

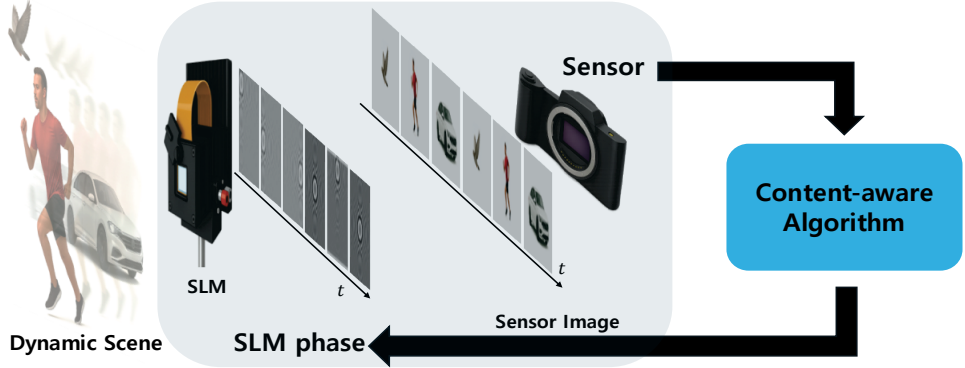

Fig. 2. System overview of the content-aware foveated camera.

Because an LCOS SLM updates the phase of each pixel by changing the applied voltage, the step response is relatively independent of the displayed spatial phase distribution. This inertia-free, pattern-agnostic update stands in sharp contrast to galvanometric scanners and other mechanically limited steering devices. Exploiting this property, we implement a time-division multiplexed, content-aware foveated camera that rapidly alternates among a sequence of user-selected sub-fields of view or ROIs. Through frame interleaving, the system supports dynamic management of multiple ROIs at the cost of reduced frame rates. This architecture enables real-time reconfiguration of both the number and spatial distribution of ROIs, allowing algorithms to continuously optimize between temporal resolution and field of view coverage.

Figure 2 illustrates the system architecture. On the SLM, at each exposure, the SLM displays a Fresnel phase pattern with focal length $f$, aperture $D$, and steering parameters $(x_f, y_f)$, as defined in Eq. 2. When the steering parameters are zero $(x_f = 0, y_f = 0)$, this reduces to the on-axis Fresnel phase described in Eq. 1; otherwise, the phase pattern steers the beam off-axis to the desired ROI location. During initialization, a set of $N$ ROIs is specified. Operation proceeds through continuous cycles, where each cycle consists of $N$ sequential frames, with each frame dedicated to monitoring a specific ROI. Within each cycle, frame $i$ (for $i$ = 1, 2, ..., $N$) is assigned to monitor ROI $i$. During frame $i$, the following sequence occurs: SLM displays the phase pattern $\phi_i(x, y; f_i, x_{f,i}, y_{f,i})$ corresponding to that ROI, the sensor captures an exposure, and a dedicated tracker $\mathcal{T}_i$ processes the captured frame to extract the target's coordinates $(x_i, y_i)$ within the ROI. The system then computes an error vector relative to the sensor center $(x_c, y_c)$ as $\left(e_i^{(x)}, e_i^{(y)}\right) = (x_i - x_c, y_i - y_c)$. This error vector provides direct feedback for a closed-loop control scheme that updates the SLM parameters for each ROI. Expressing the control law in discrete time with cycle index $n$, a proportional-derivative (PD) controller can be employed for lateral steering:

$$\Delta x_{0i}[n] = K_p^{(x)} e_i^{(x)}[n] + K_d^{(x)}\left(e_i^{(x)}[n] - e_i^{(x)}[n-1]\right). \qquad (5)$$

An analogous expression governs $\Delta y_{0i}[n]$ for vertical steering, with corresponding proportional and derivative gains $K_p^{(y)}$ and $K_d^{(y)}$. Then, the system revisits ROI $i$ in the subsequent cycle, and the SLM applies these updated parameters to generate the updated phase pattern. The SLM's stable and deterministic phase response ensures that this straightforward control approach achieves robust, real-time target tracking and foveated imaging without requiring complex model-based compensation.

## 3. Experimental Results

*3.1 Imaging setup and SLM sampling limits*

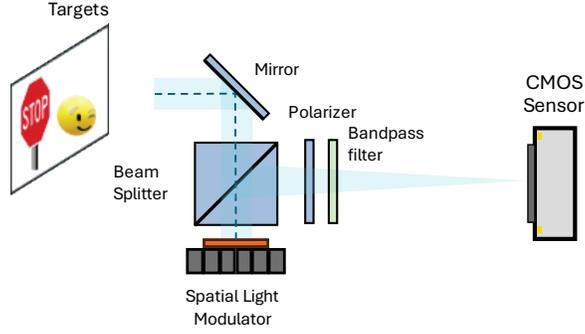

Fig. 3. Experimental setup.

To demonstrate the feasibility and potential of the content-aware foveated camera, we built a working prototype. For simplicity, we implemented a telephoto configuration in which the SLM itself serves as the imaging lens, without additional optics. The experimental setup is shown in Fig. 3. The programmable element is a reflective, phase-only SLM (HSP1920-488-800-HSP8, Meadowlark Optics, Inc.) with a resolution of 1920 × 1152 pixels and a pixel pitch of 9.2 μm, yielding an active modulation area (clear aperture) of 10.6 mm along the short side. Light from the scene is redirected by a mirror to a beamsplitter cube, modulated by the SLM, and then passes through a linear polarizer (LPVISE100-A, Thorlabs, Inc.) and a 10 nm FWHM band-pass filter centered at 550 nm (FBH550-10, Thorlabs, Inc.). After free-space propagation, the image is formed on a CMOS sensor positioned at the image plane (Sony α7 II body, lens removed).

The SLM is driven over a PCIe interface, with a measured update rate exceeding 200 Hz. In our experiments, the overall frame rate is limited by the commercial camera (< 30 Hz), so we reduced the SLM update rate by inserting inter-frame delays. Using a high-speed sCMOS sensor would allow operation closer to the SLM's native rate. We began with the on-axis diffractive lens in Eq. 1; coarse focus was set by tuning the focal length $f$ of the Fresnel phase profile loaded on the SLM. Loading Fresnel phases with different off-axis parameters $(x_f, y_f)$ translates the image at the sensor, thereby steering the region of interest (ROI).

In the demonstration, the target is 2750 mm from the SLM aperture. Based on the scene scale, we choose $f = 293.5$ mm. According to the imaging relation in Eq. 3, the SLM–sensor separation is 265.2 mm, which corresponds to an effective F-number FN $= f/D = 27.7$ (using the 10.6 mm short-edge aperture). The camera's 36 × 24 mm sensor then subtends a field width of 249 mm at the object plane, i.e., roughly ±2.6° half-angle. Consistent with the estimates in Fig. 1 (c,d), diffraction-limited quality is maintained for monochromatic operation over the designed field even with moderate off-axis steering.

A practical limitation at larger deflection is phase under-sampling on the SLM. The Fresnel phase in Eq. 1 is a chirped function, i.e., its spatial frequency increases with radius, so the maximum fringe frequency rises further when the lens is shifted off-axis. Using the paraxial approximated quadratic form of the continuous phase (Eq. 2) in polar coordinates and writing the off-axis magnitude as $r_f = \sqrt{x_f^2 + y_f^2}$, the radial fringe frequency (cycles/m) of the wrapped phase is

$$v_r = \frac{1}{2\pi}\frac{\partial \phi}{\partial r} \approx -\frac{r + r_f}{\lambda f}, \tag{6}$$

so the largest magnitude occurs at the aperture edge opposite the deflection, $r = D/2$:

$$|v_r|_{\max} \approx \frac{1}{\lambda f}\left(\frac{D}{2} + r_f\right) = \frac{1}{2\lambda \cdot \text{FN}} + \frac{|\theta|}{\lambda} \tag{7}$$

where we used $r_f = f\theta$ for small steering angle $\theta$ and $\text{FN} = f/D$. With F-number=27.7 and $\lambda = 550$ nm, the on-axis edge term is $\frac{1}{2\lambda \cdot \text{FN}} \approx 3.3 \times 10^4$ cycles/m. The SLM's 9.2 μm pixel pitch gives a Nyquist limit of $\nu_{\text{Nyq}} = \frac{1}{2p} \approx 5.4 \times 10^4$ cycles/m, so a conservative aliasing-free bound is

$$|\theta| < \lambda \nu_{\text{Nyq}} - \frac{1}{2\text{FN}} \approx 0.68°. \qquad (8)$$

Note that this theoretical aliasing-free bound is highly conservative, as it requires the aliasing-free condition to be satisfied across the entire diffractive lens. Empirically, in our setup, strong aliasing artifacts become prominent for $|\theta| > 1.7°$, so we constrain $|\theta| < 1.7°$ (i.e., $r_f < 8.7$ mm) during experiments. Using an SLM with a smaller pixel pitch would greatly increase the allowable off-axis deflection, as currently the 9.2-μm pixel size is much larger than the working wavelength of 550 nm. As SLM pixel densities continue to rise, the accessible off-axis range will expand, further broadening the applicability of this approach.

### 3.2 Single-target tracking

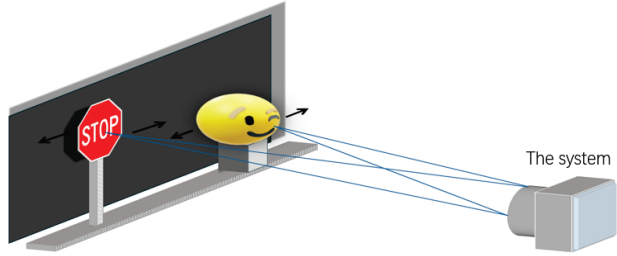

Fig. 4. Experimental setup for target tracking.

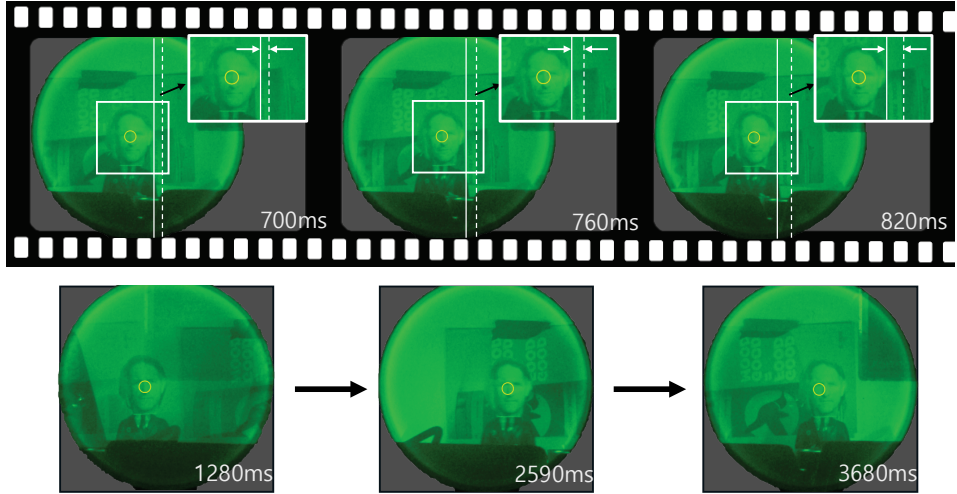

Fig. 5. Single-ROI tracking results (see Visualization 1 and Visualization 2). (a) Three consecutive frames captured by the camera at a refresh rate of 17 fps, with the stationary background marked by white dashed lines, while the moving target is marked by white solid lines. (b) Three representative key frames selected from the tracking process, showing the ROI adaptation at different time points.

As a simplified scenario, we first demonstrate the video-rate tracking capability of the system by tracking the random movement of a single target. The experimental setup is shown in Fig.4, and a generated synthetic human face object [17] is printed on a piece of paper as the single target to be tracked. To track the single target, we first initiate an on-axis Fresnel lens pattern on the SLM with a preset focal length, with the center point of the face object tracked by the Viola-Jones algorithm [18]. Next, as the single target moved, we calculated the displacement between the newly detected center point of the face on the sensor and the center of the image to generate the error vector. Then, based on the proportional-derivative control scheme described in Sec. 2, we computed the updated Fresnel lens pattern for the SLM, thereby enabling the ROI to track the center of the object.

Figure 5 shows the tracking results for the horizontal translation of the moving object. The top row displays several consecutive frames with a frame rate exceeding 17 fps. This frame rate is limited by the camera's frame rate and latency from its consumer-level video interface. White dashed lines in Fig. 5 mark the stationary background as a position reference. As the object of interest moves toward the right side of the screen, the white dash lines (marking stationary background) are seen to shift toward the right, indicating that the SLM adjusts the ROI to follow the object of interest in real time. The bottom three images of Fig. 5 show several key frames during one tracking process. The maximum deflection range is set to ±1.7°, as discussed in Sec. 3.1. The full video and more demonstrations, including tracking of circular movements, can be found in Visualization 1 and Visualization 2.

### 3.3 Multiple-target tracking

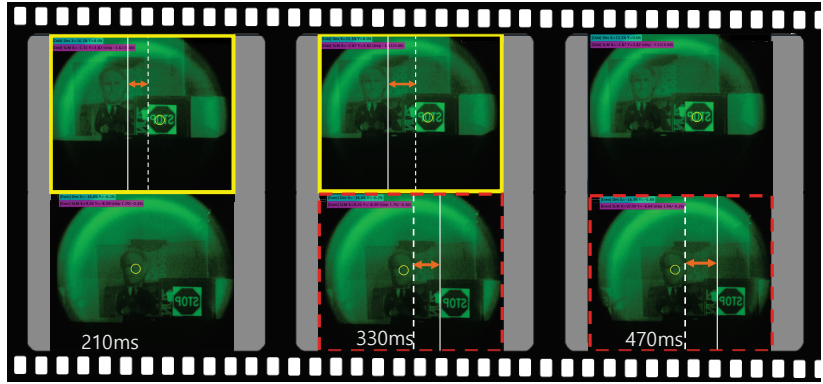

Fig. 6. Three consecutive frames from the dual-ROI tracking experiment (see Visualization 3). Frame updates are interleaved, so only one ROI is active per frame: yellow boxes denote update frames for the stop-sign ROI (top), and red boxes denote update frames for the face ROI (bottom). The dashed white line marks the tracked target; the solid line marks a background reference on a non-tracked region.

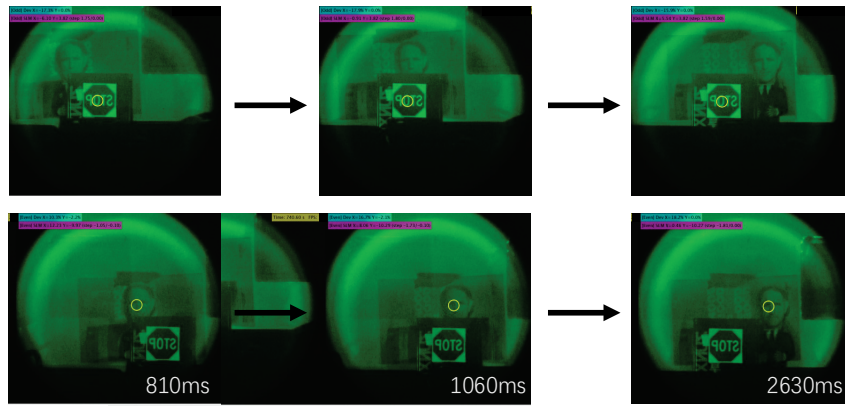

Fig. 7. Three representative key frames from the dual-ROI tracking process, showing the evolution of both ROIs at different stages (see Visualization 3).

The proposed system offers a unique capability for multiple target tracking—simultaneously tracking multiple objects through per-frame time-division multiplexing. The experimental setup remains the same as the single target tracking configuration, except that a printed stop sign is added alongside the human face object, with both objects able to move freely.

In multiple target tracking configurations, particularly in our demonstration with two targets, the even frames in the time sequence are assigned to the first object to be tracked, while the odd frames are assigned to the other object. As in single-target tracking, we run the Viola-Jones algorithm with two different cascade classifiers, each trained for a distinct object (stop sign and human face). By applying the same PD control scheme independently to odd and even frames, and alternating between the corresponding calculated Fresnel lens patterns, we can update two ROIs independently, enabling simultaneous tracking of two different objects. This scheme generalizes to $M$ targets by allocating $M$ interleaved frames per cycle (each ROI operating at the base frame rate divided by $M$).

Figure 6 presents three consecutive frames from the dual-ROI experiment. The top row tracks the stop sign, and the bottom row tracks the human face. In each frame, only one ROI is active and therefore highlighted. Yellow boxes mark update frames for the stop-sign ROI (top row), and red boxes mark update frames for the face ROI (bottom row), which are consistent with the alternate-frame scheme. The dashed white line marks the tracked object, while the solid line marks a background reference drawn on a non-tracked region (i.e., not stabilized by the controller). When the stop sign is displaced (yellow-boxed frames), the SLM keeps it centered within its ROI, which appears as a leftward shift of the background reference in the top row; meanwhile, the bottom-row ROI remains unchanged because the face is stationary in this sequence, confirming that the two control loops operate independently.

Figure 7 presents three representative key frames from the same dual-ROI tracking sequence, illustrating how both ROIs evolve across the interleaving cycle. The maximum steering deflection is again limited to $\pm 1.7°$ (Section 3.1). Because the ROIs are updated sequentially—odd/even frames mapped to different targets—each ROI is refreshed at half the base frame rate. The full video that demonstrates the real-time dynamics can be seen in Visualization 3. With an SLM refresh rate >200 Hz, a suitably matched camera would permit time-division multiplexing over many ROIs, enabling, in principle, the tracking of dozens of objects at video rate.

## 4. Conclusions

We have demonstrated a content-aware foveated imaging system that integrates a phase SLM with real-time target detection to achieve video-rate multi-object tracking while substantially reducing data throughput. The key innovation lies in combining inertia-free optical steering with content-driven ROI selection, enabling deterministic frame-interleaved sampling without sacrificing contextual awareness. This closed-loop approach represents a paradigm shift from passive, uniform acquisition to active, adaptive perception, addressing the growing data bottleneck while enabling new resource-constrained applications.

The phase SLM's ability to synthesize wavefronts at a video rate allows flexible trade-offs among steering, zoom, and aberration compensation. In microscopy, it can replace the tube lens; in cameras, it serves as either the primary lens or combines with conventional optics. The concept scales from microscopes to macroscopic cameras and extends naturally to 3D sensing through SLM-based refocusing.

Looking forward, this framework complements emerging technologies, including compressive sensing, event-driven sensors, and edge computing accelerators. By embedding the detection-tracking loop at the sensor and minimizing data movement, this approach enables applications in autonomous systems and biomedical imaging where current data rates remain prohibitive.

**Funding.** National Institutes of Health (R01HL16531, RF1NS128488, R35GM128761); Department of Energy (DE-SC0025928).

**Disclosures.** The authors declare no conflicts of interest.

**Data availability.** Data underlying the results presented in this paper are not publicly available at this time but may be obtained from the authors upon reasonable request.